\documentclass[12pt,preprint]{elsarticle}

\usepackage{graphicx}

\usepackage{amssymb}
\usepackage{amsmath}
\usepackage{amsthm}

\graphicspath{{/home/petri/Astrolib++/Champ/}}

\journal{Computer Physics Communications}

\begin{document}

\begin{frontmatter}



  \title{A spectral method in space and time to solve the
    advection-diffusion and wave equations in a bounded domain.}


\author{P\'etri, J\'er\^ome}

\address{Observatoire Astronomique de Strasbourg, Universit\'e de
  Strasbourg, CNRS, UMR 7550, 11 rue de l'Universit\'e, 67000
  Strasbourg, France.}

\begin{abstract}
  The advection-diffusion and wave equations are the fundamental
  equations governing any physical law and therefore arise in many
  areas of physics and astrophysics. For complex problems and
  geometries, only numerical simulations can give insight into
  quantitative and accurate behavior of the seeked solutions.  The
  standard numerical algorithm to solve partial differential equations
  is to split the space and time discretisation separately into
  different uncorrelated methods.  Time is usually advanced by
  explicit schemes, or, for too restrictive time steps, by implicit or
  semi-implicit algorithms.  This separate time and space slicing is
  artificial and sometimes unpractical.  Indeed, treating space and
  time directions symmetrically and simultaneously without splitting
  is highly recommended in some problems like diffusion.

  It is the purpose of this work to present a simple numerical
  algorithm to solve the standard linear scalar advection-diffusion
  and wave equations using a fully spectral method in a
  two-dimensional Cartesian $(x,t)$ bounded space-time domain.
  Generalization in three-dimensions $(x,y,t)$ (two space dimensions)
  is straightforward and shown for the pure diffusion problem. The
  basic idea is to expand the unknown function in Chebyshev
  polynomials for the spatial variables~$(x,y)$ as well as for the
  time variable~$t$.  We show typical examples and demonstrate the
  spectral accuracy of the method.  Nevertheless, we emphasize that
  the method requires to invert a large matrix that can be very time
  and memory consuming, especially for high resolution and/or higher
  dimensional problems.  This occurs already for modest resolutions in
  two-dimensions like a grid of $129\times129$ points in $(x,t)$. It
  becomes even more severe for three dimensions, we were unable to
  deal with resolution higher than $33\times33\times33$. The next step
  would be to supersede the straightforward direct inversion matrix
  operation by a more efficient and less memory demanding algorithm to
  solve large linear systems.  This is unavoidable if one wants to
  study higher spatial dimensional setups.  The aim of this study is
  to demonstrate the feasibility of such codes. Moreover, the great
  advantage of fully spectral methods resides in their high-accuracy
  for a relatively small number of grid points (for sufficiently
  smooth solutions) compared to standard time-stepping techniques.
\end{abstract}


\end{frontmatter}


\section{Introduction}
\label{}



Any physical law is represented in an abstract space called
space-time.  Physically meaningful quantities evolve in space and time
with prescribed spatial and temporal relations. Mathematically, this
is translated into a set of partial differential equations (PDEs) with
appropriate initial and boundary conditions to obtain a well-posed
problem.  They describe the fundamental behavior of any physical
system by local interactions.  Analytical solutions are very difficult
to find and are only available for a very restricted number of PDEs
and configurations.  Therefore, looking for numerical solutions to
these PDEs is an important task in order to make any prediction or
fitting observations and experiments.

Several techniques allow to study the behavior of physical systems by
looking for numerical solutions of the corresponding PDEs. Finite
difference is the most common technique and most straightforward one,
sampling space and time on a finite grid. For shocks in hydrodynamical
or MHD flows, finite volume schemes are preferred in order to
conserve, within machine accuracy, physical quantities such as total
energy and momentum if required.

Spectral or pseudo-spectral methods expand the unknown functions on a
well chosen basis of polynomials or other special functions,
depending on the geometry and on the boundary conditions.

Particle in cell simulations are common in plasma physics where
wave/particle interactions can produce some significant effects not
seen in an MHD or multi-fluid picture \citep{Birdsall2005,
  1988csup.book.....H}. They combine a set of charged particles to the
ambient electromagnetic field via Maxwell equations.

For spatially multidimensional problems, one has the freedom to choose
the most convenient method for each direction independently. For
instance in cylindrical geometry, Fourier spectral method in the
azimuthal direction, finite volume in the radial direction and finite
difference in time.

Splitting the PDE into a space and a time direction is arbitrary. From
the theory of general relativity, we know that covariance implies to
treat space and time on the same level. Such approach can reveal very
fruitful to investigate relativistic stars in general relativity.

An extensive literature on spectral method with Fourier, Chebyshev and
Legendre transforms can be found in \cite{Boyd2001} with applications
to different geometries.  Specific examples for fluid dynamics are
reported in \cite{Canuto2006, Canuto2007} for single and multi-domain
complex flows. It is well known that discontinuities, such as shocks
appearing in fluid mechanics, give rise to the Gibbs phenomenon,
degrading the convergence properties of any spectral method. This is
usually circumvent by introducing some filtering or smoothing,
applying artificial viscosity for instance. This is an ongoing active
research field \citep{2003JCoPh.192..325D}.

Spectral methods in time have been used to solve the Korteweg de Vries
and Burger equations with spatial periodic solutions, expanded in
Fourier series, \cite{1992JCoPh.102...88I}. Applications to
relativistic situations has been undertaken by Hennig and Ansorg
\cite{2009JHDE..6.161H}. Here we follow the same method, namely,
expansion of the unknown function in terms of Chebyshev polynomial
with respect to space and time coordinate and reformulation of the
initial-boundary value problem in terms of an algebraic system for the
Chebyshev coefficients or equivalently, for the function values at the
collocation points.  They solved the wave equation in Minkowski
space-time for the spherically symmetric geometry. In order to avoid
artificial boundary condition at a finite distance, they used a
compactification method. Contrary to their work, we restrict to
spatially compact domain and linear equations (solved by direct method
compared to their Newton-Raphson method allowing to treat also
non-linear problems) of first and second order.  A review on
(pseudo-)spectral methods emphasizing astrophysical applications can
be found in \cite{lrr-2009-1Grandclement}.

In this paper, we design a fully implicit spectral method to solve
PDEs in one space dimension with fixed and outgoing wave boundary
conditions, using a Cartesian coordinate system. Two spatial
dimensions are also briefly discussed. Because of the lake of
periodicity in the computational bounded domain, Chebyshev polynomial
expansion is highly recommended in both space and time directions.
Useful properties about Chebyshev polynomials are reminded in
section~\ref{sec:Chebyshev}.  Next, in section~\ref{sec:Advection} we
determine the linear advection operator to solve the full advection
equation with general boundary conditions.  The same procedure is then
applied in section~\ref{sec:Onde} for the linear wave operator in
order to solve the wave equation with outgoing or fixed boundary
conditions. Section~\ref{sec:Diffusion} depicts the diffusion equation
in two spatial dimensions. Results and tests for these simple examples
are shown in section~\ref{sec:Results} before concluding in
section~\ref{sec:Conclusion}.

\section{A reminder on Chebyshev polynomials\label{sec:Chebyshev}}

We start by recalling some important features of Chebyshev polynomials
as well as some useful properties about their derivatives.

In the most general compact computational domain, the space-time
interval is $(x,t) \in [a,b] \times [t_1,t_2]$. Without loss of
generality, it is always possible to introduce new independent
variables~$(\zeta,\tau)$ such that the scaling is recast into the
range $(\zeta,\tau) \in [-1,1] \times [-1,1]$ by a linear mapping. For
completeness, this mapping looks like
\begin{equation}
\label{eq:Mapping}
  x = \frac{b-a}{2} \, \zeta + \frac{b+a}{2}
\end{equation}
where the variable~$\zeta$ lies in the window~$[-1,1]$. The same
linear transformation would apply to the time coordinate~$t$ replaced
by~$\tau$ or any other spatially bounded variable.  We restrict our
discussion to this normalized domain.

\subsection{Definition and properties}

Chebyshev polynomials, denoted by $T_k(x)$, are defined in the
interval $x\in[-1,1]$ such that
\begin{equation}
  \label{eq:Definition_Chebyshev}
  T_k(x) = \cos(k\,\arccos(x))
\end{equation}
where $k$ is a positive integer. Their boundary values as well as
those of their first and second derivatives are easily found to be
\begin{eqnarray}
 T_k(\pm1) & = & (\pm1)^k \\
 T_k'(\pm1) & = & (\pm1)^{k+1} \, k^2 \\
 T_k''(\pm1) & = & (\pm1)^{k} \, k^2 \, \frac{k^2-1}{3}
\end{eqnarray}
These results are useful to impose special boundary conditions of
Dirichlet, Neumann or Robin type, as we will see later.

\subsection{Interpolation function}

Any well behaved function~$f(x)$, let us say continuous as well as
several of its derivatives, defined on the interval~$[-1,1]$, is
expanded in a series of Chebyshev polynomials. The interpolation of
$f$ containing $N$~terms is denoted by $f_N$ and written as
\begin{equation}
  \label{eq:Dvlpt_Chebyshev}
  f_N(x) = \sum_{k=0}^{N-1} f_k \, T_k(x)
\end{equation}
$f_k$ are the coefficients of the expansion.  Chebyshev polynomials
are closely related to the trigonometric cosines functions as seen
from their definition Eq.~(\ref{eq:Definition_Chebyshev}). The series
Eq.~(\ref{eq:Dvlpt_Chebyshev}) can therefore be computed by a Fourier
cosines transform, employing a Fast Fourier Transform (FFT) algorithm.
Because of the relation with the FFT algorithm, the first term~$f_0$
appears usually with a factor~$1/2$ \citep{1992nrca.book.....P}. Here
we suppress this normalization factor of the constant coefficient for
later convenience when dealing with the different linear operators and
matrices. We emphasize that for computational domains that are not
normalized, linear transformations such that those of
Eq.(\ref{eq:Mapping}) can be useful. The Chebyshev series expansion is
computed most efficiently and accurately with help on FFTs. The
function~$f$ is discretised on a grid corresponding to the
Gauss-Lobatto collocation points defined by
\begin{eqnarray}
\xi_i & = & \cos\left( \frac{i\,\pi}{N-1} \right)
\end{eqnarray}
with $i\in[0..N-1]$.  This links the coefficients in the Chebyshev
space to the real physical space. In the remainder of this paper, we
will assume that $(x,t)\in[-1,1]^2$ (therefore renaming $(\zeta,\tau)$
into $(x,t)$).

For PDEs, partial differentiation of the unknown quantities at any
order with respect to one coordinate, space or time, is required. We
therefore look for the relation between the coefficient of the
derivative of any order to the coefficient of the interpolant function
itself.  This is compulsory in order to solve the discretised problem
and is explained in the next paragraph.

\subsection{Derivative operator}

Differentiation of~$f$ can be interpreted as an algebraic
multiplication in the coefficient space. Indeed, let any
function~$f(x)$ possess an approximate series expansion with Chebyshev
coefficients~$f_k$ given by Eq.~(\ref{eq:Dvlpt_Chebyshev}). Those of
its first derivative~$f'$ are given by a simple matrix multiplication
\begin{equation}
  \label{eq:Derivation}
  f'_k=\sum_{i=0}^{N-1} \mathcal{D}_{ki}\,f_i
\end{equation}
with $k\in [0..N-1]$. The matrix representation of the derivative
operator in the Chebyshev coefficient space reads
\begin{equation}
  \mathcal{D}_{ik} = 
  \begin{pmatrix}
    0 & 1 & 0 & 3 & 0 & \ldots & k & \ldots & N-1 \\
    0 & 0 & 4 & 0 & 8 & & 0 & \ldots & 0 \\
    0 & 0 & 0 & 6 & 0 & & 2\,k & \ldots & 2\,(N-1) \\
    0 & 0 & 0 & 0 & 8 & & 0 & \ldots & 0 \\
    \vdots & 0 & 0 & 0 & 0 & \ddots & \vdots & & \vdots\\
    \vdots & 0 & 0 & 0 & 0 & \ddots & 2\,k & & 0 \\
    \vdots &  &  &  &  & \ldots & 0 & & 2\,(N-1) \\
    0 &  &  &  &  & \ldots & & & 0
  \end{pmatrix}
\end{equation}
It is an upper triangular matrix with its diagonal terms equal to
zero.

The same procedure can be repeated for higher order derivatives of
order~$n$ denoted by $f^{(n)}$. We need only to take the corresponding
power of the matrix~$\mathcal{D}_{ik}$ such that the coefficient of
$f^{(n)}$ are
\begin{equation}
  \label{eq:Derivation2}
  f^{(n)}_k = \sum_{i=0}^{N-1} (\mathcal{D}_{ki})^n\,f_i
\end{equation}
We summarize the results by writing the expansion of the $n$-th
derivative of the interpolant as
\begin{equation}
\label{eq:Dvlpt_Derivee_Chebyshev}
f_N^{(n)}(x) = \sum_{i=0}^{N-1} \sum_{k=0}^{N-1} (\mathcal{D}_{ki})^n \, f_i \, T_k(x)
\end{equation}

\subsection{Product}

For linear differential operators such as the Laplacian or for an
arbitrary linear combination of partial derivatives, it is helpful to
find a matrix expression of that operator in the coefficient space.

In order to work out this operator explicitly, we need to relate the
coefficients of a product of two functions to those of these two
functions.  Let us be two functions $f$ and $g$ defined in $[-1,1]$.
Given their respective expansion coefficients $f_k$ and $g_k$ for
their interpolant $f_N$ and $g_N$, we look for the expression of the
coefficient of their product $f_N\,g_N$.  These are found by
introducing another ``matrix'' possessing three indices and denoted
by~$\mathcal{P}_{ikl}$, with $(i,k,l)\in[0..N-1]^3$, such that
\begin{equation}
  \label{eq:Produit}
  (f_N \, g_N)_i = \sum_{k=0}^{N-1} \sum_{l=0}^{N-1} \mathcal{P}_{ikl} \, f_k \, g_l
\end{equation}
For concreteness, the first elements of this matrix are reported
explicitly in the following lines. The first ``sub-matrix'' is diagonal and reads
\begin{equation}
\mathcal{P}_{0kl} = 
\begin{pmatrix}
1      &      0 & \ldots &        & 0 \\
0      &    1/2 & \ddots &        & \vdots \\
\vdots & \ddots &    1/2 & \ddots & \vdots \\
\vdots &        & \ddots & \ddots & 0 \\
0      & \ldots & \ldots & 0      & 1/2
\end{pmatrix}
\end{equation}
The second submatrix has a vanishing diagonal such that
\begin{equation}
\mathcal{P}_{1kl} = 
\begin{pmatrix}
0      & 1      & 0      & \ldots & 0 \\
1      & 0      & 1/2    & \ddots & \vdots \\
0      & 1/2    & \ddots & \ddots & 0\\
\vdots & \ddots & \ddots & \ddots & 1/2 \\
0      & \ldots & 0      & 1/2    & 0
\end{pmatrix}
\end{equation}
and, finally, the third submatrix is
\begin{equation}
\mathcal{P}_{2kl} = 
\begin{pmatrix}
0      & 0      & 1      & 0      & \ldots & 0 \\
0      & 1/2    & 0      & 1/2    & \ddots & \vdots \\
1      & 0      & 0      & \ddots & \ddots & 0 \\
0      & 1/2    & \ddots & \ddots & \ddots & 1/2 \\
\vdots & \ddots & \ddots & \ddots & 0      & 0 \\
0      & \ldots & 0      & 1/2    & 0      & 0
\end{pmatrix}
\end{equation}
The remaining submatrices for $i>2$ can be deduced easily. Note that
all these submatrices are symmetric with respect to the last two
indices $k$ and~$l$ owing to the symmetry of the product of two
functions, i.e.  $f_N\,g_N = g_N\,f_N$.

With the derivatives and the multiplication defined as matrices given
above, any linear differential operator in space and time can be
written in any number of spatial dimensions. For instance, the operators
\begin{equation}
  \label{eq:OpDiff1}
  \frac{1}{x} \, \frac{\partial}{\partial x}
\end{equation}
and
\begin{equation}
  \label{eq:OpDiff2}
  \frac{1}{x^2} \, \frac{\partial^2}{\partial x^2}
\end{equation}
are encountered when expressing the Laplacian in cylindrical and
spherical geometry. This is very useful for deriving the matrix
operator according to such linear differential operator.

In the following sections, two straightforward applications are shown
for the advection and wave equations. They serve as a starting point
to check the numerical algorithm and look for the convergence
properties before shifting to more advanced schemes dealing with
several geometries and several unknown functions. However, these
complications are left for future work. Nevertheless, the two space
dimensions set up is discussed in the constant coefficient diffusion
problem.

\section{The 1D advection equation\label{sec:Advection}}

The advection equation at constant speed is the standard benchmark
problem to test a code solving PDE equations. The exact analytical
solution is known, allowing an investigation of the accuracy of the
scheme. After a brief sketch of this equation, we develop a fully
spectral method to solve the linear advection equation.

\subsection{The general problem}

The one dimensional advection equation with appropriate boundary and
initial conditions, assuming a constant and positive speed $v>0$ and a
real parameter~$b$, is given by
\begin{eqnarray}
\label{eq:Advection}
 \frac{\partial u}{\partial t} + v \, \frac{\partial u}{\partial x} & = & b \, u + f \\
 u(x,-1) & = & u_0(x) \\
 u(-1,t) & = & B_G(t)
\end{eqnarray}
Note that compatibility between spatial and temporal boundary
conditions requires $B_G(-1) = u_0(-1)$.  We look for
solutions~$u(x,t)$ for spacetime coordinates $(x,t)\in[-1,1]^2$,
initial condition $u_0(x)$ and left boundary condition $B_G(t)$. The
exact analytical solution is known. It corresponds to a propagation
along the characteristic defined by $dt/ds=1$ and $dx/ds=v$ and
satisfies $du/ds=du/dt=b\,u$. Explicitly, the solution reads
\begin{equation}
  \label{eq:SolutionAdvection}
  u(x,t) = u_0(x-v\,(t-t_0))\,e^{b\,t} + B_G\left(t-\frac{x-x_0}{v}\right) \, H(v\,(t-t_0)-(x-x_0))
\end{equation}
where $H$ is the Heaviside unit step function and $x_0 = t_0 = -1$.  Physically, the
signal~$u_0(x)$ travels along the positive $x$-axis at speed~$v$
(compatible with the left boundary condition~$B_G(t)$) with an
exponential decrease (resp.  increase) in time according to $e^{b\,t}$
for $b>0$ (resp. $b<0$).

\subsection{Spectral method in space and time}

Approximate solutions are found by expanding the unknown
function~$u(x,t)$ as a double Chebyshev series in both space and time.
We seek solutions of the two variables $(x,t)\in[-1,1]^2$ such that
\begin{equation}
  \label{eq:Dvlpt_Chebyshev_2D}
  u_N(x,t) = \sum_{i=0}^{N_x-1} \sum_{k=0}^{N_t-1} u_i^k \, T_i(x) \, T_k(t)
\end{equation}
The number of expansion coefficients in space is labeled~$N_x$ whereas
the one in time is labeled~$N_t$. Obviously, they are not restricted
to be equal.

The spatial and temporal first order derivatives of this expansion are
obtained from the matrix operator~$\mathcal{D}_{kl}^{(x)}$
and~$\mathcal{D}_{kl}^{(t)}$, the superscripts~$x$ and~$t$ label the
differentiation against spatial and temporal coordinates respectively.
Note that these matrices are squared but their dimensions are not
necessarily the same, resp.~$N_x$ and~$N_t$. The time and space
derivatives are given respectively by
\begin{eqnarray}
  \label{eq:Dvlpt_Derivee_Chebyshev_2D}
  \frac{\partial u_N}{\partial t}(x,t) & = & 
  \sum_{i=0}^{N_x-1} \sum_{k=0}^{N_t-1} \sum_{l=0}^{N_t-1}  
  \mathcal{D}_{kl}^{(t)} \, u_i^l \, T_i(x) \, T_k(t) \\
  \frac{\partial u_N}{\partial x}(x,t) & = & 
  \sum_{i=0}^{N_x-1} \sum_{k=0}^{N_t-1} \sum_{j=0}^{N_x-1}  
  \mathcal{D}_{ij}^{(x)} \, u_j^k \, T_i(x) \, T_k(t) 
\end{eqnarray}
From this, we deduce the full expression of the linear operator for
the advection equation, being
\begin{eqnarray}
  & & \frac{\partial u_N}{\partial t}(x,t) + v \, \frac{\partial
    u_N}{\partial x}(x,t) - b \, u_N(x,t) \\
  & = & \sum_{i=0}^{N_x-1} \sum_{k=0}^{N_t-1} 
  \left[ \sum_{l=0}^{N_t-1} \mathcal{D}_{kl}^{(t)} \, u_i^l +
    v \, \sum_{j=0}^{N_x-1} \mathcal{D}_{ij}^{(x)} \, u_j^k - b \,
    u_i^k \right] \, T_i(x) \, T_k(t) \\
  & = & \sum_{i,j=0}^{N_x-1} \sum_{k,l=0}^{N_t-1} \left[ \mathcal{D}_{kl}^{(t)} \, \delta_{ij} +
    v \, \mathcal{D}_{ij}^{(x)} \, \delta_{kl} - b \, \delta_{ij} \, \delta_{kl} \right] \, u_j^l \, T_i(x) \, T_k(t)
\end{eqnarray}
If present, we expand the source term similarly
\begin{equation}
  \label{eq:SourceAdvection}
  f_N(x,t) = \sum_{i=0}^{N_x-1} \sum_{k=0}^{N_t-1} f_i^k \, T_i(x) \, T_k(t)
\end{equation}
The unknowns $u_i^k$ have to satisfy the advection equation in the
interior of the integration domain. It is convenient to introduce the
four dimensional matrix defined by
\begin{equation}
 \label{eq:Tenseur4DAdvection}
 \mathcal{A}_{ijkl} = \mathcal{D}_{kl}^{(t)} \, \delta_{ij} + 
 v \, \mathcal{D}_{ij}^{(x)} \, \delta_{kl} - b \, \delta_{ij} \, \delta_{kl}
\end{equation}
such that the four dimensional system to solve reads
\begin{equation}
\label{eq:SysMat}
\sum_{j=0}^{N_x-1} \sum_{l=0}^{N_t-1} \mathcal{A}_{ijkl} \, u_j^l = f_i^k
\end{equation}
The two indices $i$ and $k$ run from 0 to $N_x-1$ and from 0 to
$N_t-1$ respectively. This does not yet include the boundary and
initial conditions.

Indeed, in order to be able to impose adequate initial and boundary
conditions, we use the penalty method, stating that
\begin{eqnarray}
  \label{eq:Penalite}
  \frac{\partial u_N}{\partial t}(-1,t) + v \, \frac{\partial
    u_N}{\partial x}(-1,t) + \tau \, v \, ( u_N(-1,t) - B_G(t) ) & = & 0 \\
  \frac{\partial u_N}{\partial t}(x,-1) + v \, \frac{\partial
    u_N}{\partial x}(x,-1) + \tau \, ( u_N(x,-1) - u_0(x) ) & = & 0
\end{eqnarray}
$\tau$ is a parameter that must be large enough to ensure stability of
the scheme. Typically we took $\tau=N(N+1)/2$.  Thus we have to add
the appropriate terms in the matrix~$\mathcal{A}_{ijkl}$ and
right-hand-side~$ f_i^k$. Noting that
\begin{eqnarray}
  u(x,-1) & = & \sum_{i=0}^{N_x-1} \sum_{k=0}^{N_t-1} u_i^k \, T_i(x)
  \, T_k(-1) \\
  & = & \sum_{i=0}^{N_x-1} \sum_{k=0}^{N_t-1} (-1)^k \, u_i^k \, T_i(x)
\end{eqnarray}
we need to include terms like $(-1)^k \, u_i^k$ into the matrix
elements~$\mathcal{A}_{ijkl}$.  Similarly, for the initial profile, we
get in terms of its expansion coefficients
\begin{equation}
  \label{eq:Expan}
  u_0(x) = \sum_{i=0}^{N_x-1} u_{0i} \, T_i(x)  \; i \in [0..N_x-1]  
\end{equation}
Thus, the quantities $u_{0i}$ should be added to the right-hand-side
element $f_i^k$. Following the same lines, the boundary condition is
expressed as
\begin{eqnarray}
  u(-1,t) & = & \sum_{i=0}^{N_x-1} \sum_{k=0}^{N_t-1} u_i^k \, T_i(-1)
  \, T_k(t) \\
  & = & \sum_{i=0}^{N_x-1} \sum_{k=0}^{N_t-1} (-1)^i \, u_i^k \, T_k(t)
\end{eqnarray}
Projecting the boundary function~$B_G(t)$ on the polynomial basis
gives
\begin{equation}
  \label{eq:ProjAdvec}
  B_G(t) = \sum_{k=0}^{N_t-1} B_{Gk} \, T_k(t) \; k \in [0..N_t-1]  
\end{equation}
These terms $(-1)^i \, u_i^k $ and $B_{Gk}$ must also be appropriately included
in the matrices~$\mathcal{A}_{ijkl}$ and~$ f_i^k$. Note the similarity
between initial Eq.~(\ref{eq:Expan}) and boundary
Eq.~(\ref{eq:ProjAdvec}) conditions. Space and time are treated
equally.

Finally, Eq.~(\ref{eq:SysMat}) is converted into a classical two
dimensional linear algebra problem by a change of variables. For the
unknown vector containing all the unknown quantities $u_i^k$, we write
\begin{equation}
  \label{eq:VecteurAdvection}
  z_s = u_j^l 
\end{equation}
with $s=j+l\,N_x$. The matrix is changed into
\begin{equation}
  \label{eq:MatriceAdv}
  A_{rs} = \mathcal{A}_{ijkl}
\end{equation}
with $r=i+k\,N_x$ and the right hand side into
\begin{equation}
  \label{eq:SourceVecAdv}
  h_r = f_i^k
\end{equation}
Finally, the system is rewritten into
\begin{equation}
  \label{eq:SystemeAdv}
  \sum_{s=0}^{N_xN_t-1} A_{rs} \, z_s = h_r  
\end{equation}
with $r\in[0..N_xN_t-1]$. 


This completes the translation of the PDE into a linear system for the
set of coefficients~$u_i^k$. It is a well posed problem, invertible
with standard linear algebra techniques.


\section{The wave equation\label{sec:Onde}}

\subsection{Basics}

Another important equation arising very often in physical systems is
the wave equation. In Cartesian coordinates, supplemented with
appropriate initial conditions, it reads
\begin{eqnarray}
  \frac{\partial^2 u}{\partial t^2}(x,t) -
  v^2 \, \frac{\partial^2 u}{\partial x^2}(x,t) & = & f(x,t) \\
  u(x,-1) & = & u_0(x) \\
  \frac{\partial u}{\partial t}(x,-1) & = & v_0(x)
\end{eqnarray}
Initial profile $u_0(x)$ and first time derivative~$v_0(x)$ must be
given.  The boundary conditions can be outgoing waves on both sides
\begin{equation}
  \label{eq:CL}
  \frac{\partial u}{\partial t}(\pm1,t) \pm v \, \frac{\partial u}{\partial x}(\pm1,t) = 0
\end{equation}
or fixed boundaries, vanishing for instance on both sides
\begin{equation}
  \label{eq:CL2}
  u(\pm1,t) = 0
\end{equation}
or a mix between both types, outgoing on one side and fixed on the other side.

\subsection{Spectral method in space and time}

The method follows exactly the same lines as for the advection
problem. The unknown function is expanded according to
Eq.(\ref{eq:Dvlpt_Chebyshev_2D}).

The full linear operator representing the wave equation becomes
\begin{eqnarray}
  & & \frac{\partial^2 u_N}{\partial t^2}(x,t) - v^2 \, \frac{\partial^2
    u_N}{\partial x^2}(x,t) \\
  & = & \sum_{i=0}^{N_x-1} \sum_{k=0}^{N_t-1} 
  \left[ \sum_{l=0}^{N_t-1} (\mathcal{D}_{kl}^{(t)})^2 \, u_i^l -
    v^2 \, \sum_{j=0}^{N_x-1}  (\mathcal{D}_{ij}^{(x)})^2 \, u_j^k \right] \, T_i(x) \, T_k(t) \\
  & = & \sum_{i,j=0}^{N_x-1} \sum_{k,l=0}^{N_t-1} \left[ (\mathcal{D}_{kl}^{(t)})^2 \, \delta_{ij} -
    v^2 \, (\mathcal{D}_{ij}^{(x)})^2 \, \delta_{kl} \right] \, u_j^l \, T_i(x) \, T_k(t)
\end{eqnarray}
We expand the source term similarly
\begin{equation}
  \label{eq:SourceOnde}
  f_N(x,t) = \sum_{i=0}^{N_x-1} \sum_{k=0}^{N_t-1} f_i^k \, T_i(x) \, T_k(t)
\end{equation}
The unknowns~$u_i^k$ have this time to satisfy the wave equation in
the interior of the integration domain with special care for
imposing the boundary and initial conditions. It is convenient to
introduce another four dimensional matrix defined by
\begin{equation}
 \label{eq:Tenseur4DOnde}
 \mathcal{O}_{ijkl} = (\mathcal{D}_{kl}^{(t)})^2 \, \delta_{ij} -
    v^2 \, (\mathcal{D}_{ij}^{(x)})^2 \, \delta_{kl}
\end{equation}
such that the four dimensional system to solve reads
\begin{equation}
 \label{eq:Tenseur2DOnde}
  \sum_{j=0}^{N_x-1} \sum_{l=0}^{N_t-1} \mathcal{O}_{ijkl} \, u_j^l = f_i^k
\end{equation}
Here also, the two indices $i$ and $k$ run from 0 to $N_x-1$ and from
0 to $N_t-1$ respectively.

The matrix is completed by imposing initial conditions. For the
unknown function, it reads
\begin{eqnarray}
  u_N(x,-1) & = & \sum_{i=0}^{N_x-1} \sum_{k=0}^{N_t-1} u_i^k \, T_i(x)
  \, T_k(-1) \\
  & = & \sum_{i=0}^{N_x-1} \sum_{k=0}^{N_t-1} (-1)^k \, u_i^k \, T_i(x)
\end{eqnarray}
This leads to a linear system
\begin{equation}
  \label{eq:SysLinOnde}
  u_0(x) = \sum_{i=0}^{N_x-1} u_{0i} \, T_i(x) \; i \in [0..N_x-1] 
\end{equation}
For its first time derivative, we have
\begin{eqnarray}
  \frac{\partial u_n}{\partial t}(x,-1) & = & \sum_{i=0}^{N_x-1}
  \sum_{k=0}^{N_t-1} u_i^k \, T_i(x) \, T_k'(-1) \\
  & = & \sum_{i=0}^{N_x-1} \sum_{k=0}^{N_t-1} (-1)^{k+1} \, k^2 \, u_i^k \, T_i(x)
\end{eqnarray}
Next, using the boundary values for the Chebyshev first derivatives,
we are led to another linear system
\begin{equation}
  \label{eq:SysLinOnde2}
  v_0(x) = \sum_{i=0}^{N_x-1} v_{0i} \, T_i(x) \; i \in [0..N_x-1]
\end{equation}

Let us finish to describe the algorithm by implementing the boundary
conditions. On one side, for outgoing wave conditions on both ends, at
$x=+1$ and $x=-1$, they respectively translate into
\begin{eqnarray}
 \sum_{j=0}^{N_x-1} \sum_{l=0}^{N_t-1} 
 \left[ \mathcal{D}_{kl}^{(t)} + v \, j^2 \, \delta_{kl} \right] \, u_j^l & = & 0 \\
 \sum_{j=0}^{N_x-1} \sum_{l=0}^{N_t-1} 
 (-1)^j \, \left[ \mathcal{D}_{kl}^{(t)} + v \, j^2 \, \delta_{kl} \right] \, u_j^l & = & 0
\end{eqnarray}
for $k\in[0..N_t-1]$.  On the other side, for fixed and vanishing boundary
conditions on both ends, it reduces to
\begin{eqnarray}
  \sum_{j=0}^{N_x-1} \, u_j^l & = & 0 \\
  \sum_{j=0}^{N_x-1} (-1)^j \, u_j^l & = & 0
\end{eqnarray}
for $l\in[0..N_t-1]$. This gives another set of $2N_t$~equations for
the matrix $\mathcal{O}$.


Eq.~(\ref{eq:Tenseur2DOnde}) is converted into another classical two
dimensional linear algebra problem by the same change of variables
except that the matrix is now
\begin{equation}
  \label{eq:MatriceOnde}
  O_{rs} = \mathcal{O}_{ijkl}
\end{equation}
and the new system
\begin{equation}
  \label{eq:SystemeOnde}
  O_{rs} \, z_s = h_r  
\end{equation}
$z_s$ and $h_r$ being already defined for the advection problem. The
same definition applies here except for the indices limits.

\section{The 2D diffusion equation\label{sec:Diffusion}}

Solving the diffusion equation requires usually a very small explicit
finite difference time stepping, preventing from a large time span
simulation. Therefore, this part of a PDE is solved by some implicit
time schemes, like the seconder order Crank-Nicholson algorithm.
Infinite spectral order of accuracy can be recovered by performing a
spectral expansion in time as shown in this paragraph. Moreover, it
applies to any number of spatial dimensions, we demonstrate the
feasibility for two-dimensional space configurations.

Let us consider the diffusion equation in the domain
$(x,y)\in[-1,1]^2$ for $t\in[-1,1]$ such that
\begin{eqnarray}
  \label{eq:Diffusion}
  \frac{\partial u}{\partial t}(x,y,t) & = & 
  \eta(x,y,t) \, \left[ \frac{\partial^2 u}{\partial x^2}(x,y,t) +
    \frac{\partial^2 u}{\partial y^2}(x,y,t) \right] \\
  u(x,y,-1) & = & u_0(x,y)
\end{eqnarray}
The initial profile $u_0(x,y)$ must be specified and $\eta$ is the
diffusion coefficient. We enforce Dirichlet boundary conditions on the
border of the domain such that
\begin{equation}
  \label{eq:CL3}
  u(\pm1,\pm1,t) = 0
\end{equation}
Imposing other boundary conditions, like Neumann or Robin type is
straightforward.  We use the Chebyshev collocation method to solve the
homogeneous boundary conditions by introducing the basis of Chebyshev
polynomials $T_k(x)$. Our choice is guided by a possible extension of
this algorithm to non-linear problems; it is easier to work in
physical space compared to the coefficient space, especially for
products of the non-linear terms and non constant diffusion
coefficient or other spatially varying characteristics of the problem
under investigation.

The homogeneous boundary condition are imposed by a Chebyshev
collocation method to get the linear system satisfied by the
coefficients of the interpolant $u_N$ given by an expansion like
\begin{equation}
  \label{eq:DvltpDiffusion}
  u_N(x,y,t) = \sum_{i=0}^{N_x} \sum_{j=0}^{N_y} \sum_{k=0}^{N_t}
  u_{ijk} \, T_i(x) \, T_j(y) \, T_k(t)
\end{equation}

The spatial and temporal derivatives of the interpolant~$u_N$ are
immediately obtained from those of $\phi_{i}(x)$. 

For the points of discretisation, we get a linear system for the
unknown coefficients~$u_{ijk}$ as
\begin{equation}
  \label{eq:DiffLin}
  L \, u = g
\end{equation}
where the square matrix~$L$ is obtained following the indexing method
proposed by \cite{Boyd2001} as
\begin{eqnarray}
  \label{eq:defL}
  L_{rs} & = & T_i(x_l) \, T_j(y_m) \, T'_k(t_n) - \\
  & & \eta(x_l,y_m,t_n) \, \left[ T''_i(x_l) \, T_j(y_m) \,
    T_k(t_n) + T_i(x_l) \, T''_j(y_m) \, T_k(t_n) \right] \nonumber
\end{eqnarray}
We do not give explicit expressions for the indices~$(r,s)$, the
collapsing of $(i,j,l)$ being clearly exposed in \cite{Boyd2001} for
2D problems. Extension to our 3D problem is straightforward. Note also
that the right-hand-side vector~$g$ in Eq.~(\ref{eq:DiffLin}) contains
the initial~$u_0(x_l,y_m)$ as well as the boundary conditions.

Now we have all the tools necessary to check and test the code,
especially the treatment of the boundary conditions. We discuss the
results in the next sections.

\section{Check and test of the algorithm\label{sec:Results}}

We now give some typical applications to check our algorithm on simple
initial value problems like advection and wave propagation.  Different
boundary conditions are also investigated for the wave equation.

\subsection{Advection equation}

Our advection equation at constant positive speed propagates a signal
along the positive $x$-axis. For the initial value, we take either a
Gaussian shape vanishing at the boundaries $x=\pm1$ (in order to
remain consistent with the boundary conditions) like
\begin{equation}
  \label{eq:Initial}
  u_0(x) = ( 1 - x^2 ) \, e^{-10\,x^2}
\end{equation}
or simply zero $u_0(x)=0$ but with incoming signal from the left or a
source term. Actually we tried this prescription and could not get
errors less than $10^{-7}$ (see definition of the error below) because
the prescribed initial profile is not infinitely smooth at the
boundary $\pm1$. We therefore tried a more smooth function at these
boundaries such that they are satisfied within numerical accuracy. The
basic idea is to use a bunch of shifted Gaussians such that
\begin{equation}
  \label{eq:InitialSmooth}
  u_0(x) = e^{-10\,x^2} - e^{-10\,(x-2)^2} - e^{-10\,(x+2)^2} 
  + e^{-10\,(x-4)^2} + e^{-10\,(x+4)^2} 
\end{equation}
This way, $u_0$ remains very close to zero and behaves smoothly, see
for instance \cite{PhysRevD.75.024006}.

We performed different sets of run. In the first set, the choice of a
Gaussian is useful to look for outgoing waves, fixing the left
boundary to zero. In a second set, the vanishing initial condition is
set up in order to test in-going left boundary conditions. For instance
we took a incoming signal such that for $x=-1$
\begin{equation}
  \label{eq:Boundary}
  B_G(t) = \sin(2\,\pi\,t)
\end{equation}
This is applied on the left side at any time $t\in[-1,1]$. The
particular choice of a frequency of $\omega=2\,\pi$ ensures to get
exactly one period inside the simulation box at the final time of
simulation. Any other value would also be fine.

Finally, in a last set of runs, a source term has been add, prescribed
as a harmonic function in time and Gaussian shape in space as follows
\begin{equation}
  \label{eq:Source}
  s(x,t) = \frac{\sin(2\,\pi\,t)}{2\,\pi} \, ( 1 - x^2 ) \, e^{-20\,x^2}
\end{equation}
In order to avoid pollution of the solution by inhomogeneous boundary
conditions, in this case, we force again the left boundary to zero.

\begin{figure}
  \centering
  \includegraphics[width=0.45\textwidth]{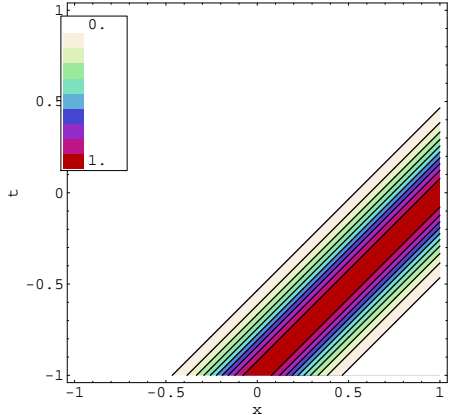}
  \caption{Solution of the advection equation for the initial Gaussian
    function and no in-going information, $B_G(t)=0$. The speed is set to
  unity, $v=1$. No spurious effects appear on the boundaries.}
\label{fig:Advection1}
\end{figure}

Let us have a look on some results for each set of runs.  The Gaussian
signal travelling along the positive $x$-axis is depicted in a
space-time plot in Fig.~\ref{fig:Advection1} with a constant speed
$v=1$. There is no incoming information from the left side $x=-1$, it
is imposed to zero therefore remaining at these value during the full
time span. As expected for these normalization constants, the
characteristic curves are inclined to $45^o$ with respect to the time
axis. The signal leaves the simulation box on the right side $x=1$
without any spurious oscillation or other undesirable effects.  The
numerical implementation of the outgoing wave condition is perfectly
reproduced by our algorithm, to machine accuracy.

Moreover, the convergence properties are checked by inspecting the
behavior of the error against the number of collocation points. We
define this error by
\begin{equation}
  \label{eq:Erreur}
  \epsilon = \textrm{max}|u_{\rm an}(x_i,t_k) - u_{\rm num}(x_i,t_k)|_{(i,k)\in[0..N-1]^2}
\end{equation}
where $u_{\rm an}$ corresponds to the exact analytical solution
Eq.~(\ref{eq:SolutionAdvection}) and $u_{\rm num}$ to the numerical
spectral approximation.

The trend is summarized in tab.~\ref{tab:AdvectionErreur1}.  The
minimum relative error falls down to less than~$10^{-10}$. This is
achieved for a number of discretisation points $N_x=N_t=129$, the
maximum we can reach on our computer.
\begin{table}
  \centering
  \begin{tabular}{ccc}
    \hline
    $N_x$ & $N_t$ & error \\
    \hline
    \hline
    5 & 5 & 1.221E-01 \\
    9 & 9 & 1.044E-01 \\
    17 & 17 & 1.661E-02 \\
    33 & 33 & 6.218E-06 \\
    65 & 65 & 1.668E-10 \\
    129 & 129 & 4.799E-11 \\
    \hline
  \end{tabular}
  \caption{Convergence properties for the advection equation. The
    error is plotted against the number of discretisation points,
    being the same in~$x$ and~$t$ coordinates.}
  \label{tab:AdvectionErreur1}
\end{table}

\begin{figure}
  \centering
  \includegraphics[width=0.45\textwidth]{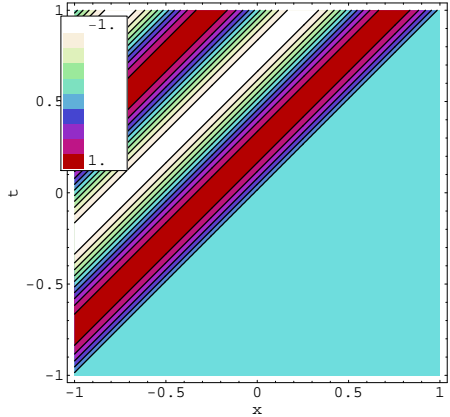}
  \caption{Solution of the advection equation with incoming left
    boundary condition. The imposed sinusoidal wave propagates to the
    right with speed $v=1$. No spurious effects appear on the
    boundaries.}
\label{fig:Advection2}
\end{figure}

Next we consider the second set of runs where homogeneous initial
conditions $u_0(x)=0$ are employed supplemented with left boundary
incoming signal prescribed by Eq.~(\ref{eq:Boundary}).  The
corresponding space-time plot is shown in Fig.~\ref{fig:Advection2}.
As expected, the imposed sinusoidal wave propagates to the right with
constant speed $v=1$. No spurious effects appear on any boundary.
At the final time of the simulation, the solution matches the function
$\sin(2\,\pi\,x)$ containing one period of the trigonometric $\sin$
function as would be expected from the exact analytical solution.

\begin{figure}
  \centering
  \includegraphics[width=0.45\textwidth]{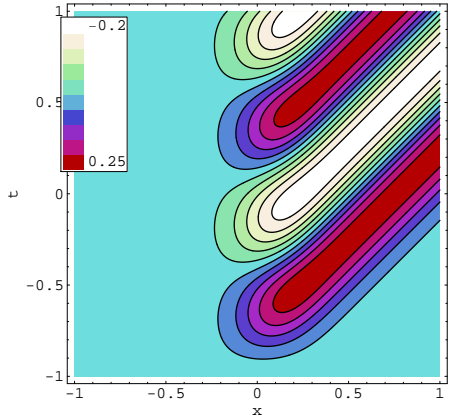}
  \caption{Solution of the advection equation with a source term. Here
    also, no spurious effects appear on the boundaries.}
\label{fig:Advection3}
\end{figure}

Finally, for the last set of runs, we still start with homogeneous
initial conditions $u_0(x)=0$ and no left boundary incoming signal,
but we add a source term~$f$ on the right hand side of the advection
equation with Gaussian shape. Results are presented in
Fig.~\ref{fig:Advection3}. Perturbations from the source arise mostly
in the middle of the box and are transported to the right along the
positive $x$-axis and leave the computational domain very smoothly in
accordance with the analytical expected solution. 

This concludes the advection part. In this section, we demonstrated
that a fully spectral method in both space and time can be used to
solve partial differential equations like the advection problem. No
particular problem arises to treat properly the boundary conditions or
the source term. Outgoing waves are handled to machine accuracy.

In the following paragraph, we show that the same conclusions apply
well to the wave equation.

\subsection{Wave equation}

For the wave equation, we use the same initial conditions as those
previously explained. However, supplementary initial and boundary
value conditions must be imposed because it is second order in both
space and time. Thus, the first order time derivative must be given. We
took simply
\begin{equation}
  \label{eq:InitialDerivee}
  \frac{\partial}{\partial t} u_0(x) = 0
\end{equation}
with additional boundary conditions on the left and right ends.

Fixed left and right boundary conditions are also interesting in order
to check the perfect reflection of the wave on both sides of the
computational domain.  We discuss the results in detail now.

\begin{figure}
  \centering
  \includegraphics[width=0.45\textwidth]{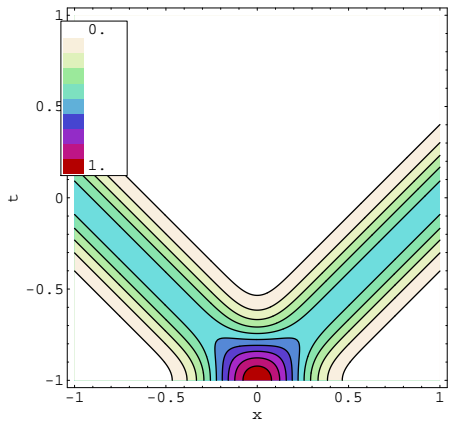}
  \caption{Solution of the wave equation with outgoing wave conditions
    and an initial Gaussian profile. Two waves emerge propagating in
    opposite direction with speed $v=1$. They leave the simulation box
    perfectly, no spurious effects.}
\label{fig:Onde1}
\end{figure}

In Fig.~\ref{fig:Onde1} the space-time plot of the solution to the
outgoing wave problem is presented for the initial Gaussian shape. As
expected, it splits into to equal parts, one propagating to the left
at speed $v=1$ and the other to right at the same speed. When reaching
the boundaries, they leave the integration domain, and the box remains
empty with $u(x,t)=0$ after a while. Boundary conditions are correctly
implemented, no oscillations or divergence are reported.

\begin{figure}
  \centering
  \includegraphics[width=0.45\textwidth]{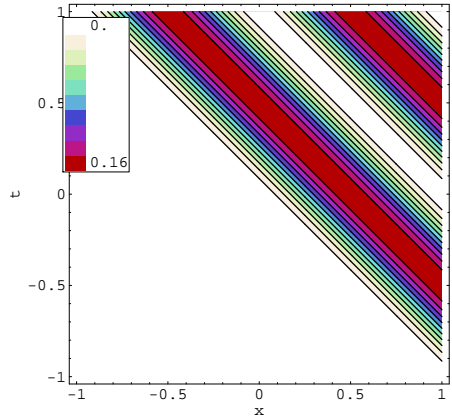}
  \caption{Solution of the wave equation with initial homogeneous and
    vanishing conditions but right incoming sinusoidal wave.}
\label{fig:Onde2}
\end{figure}

\begin{figure}
  \centering
  \includegraphics[width=0.45\textwidth]{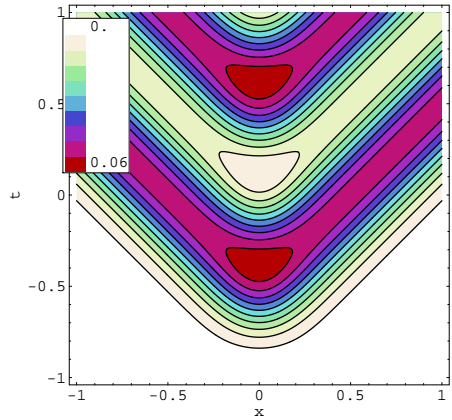}
  \caption{Solution of the wave equation with initial homogeneous and
    vanishing conditions. A source term is added and outgoing wave
    boundaries are imposed.}
\label{fig:Onde3}
\end{figure}

Next we consider an incoming wave from the right side, with sinusoidal
temporal variations, the same as for advection and outgoing left
boundary. This sinusoidal shape fills the entire box at the end of
simulation. Fig.~\ref{fig:Onde2} summarizes the trend, the
characteristic curves are inclined to $45^o$ in accordance with $v=1$.
Instead of an incoming wave, we replace it next by a source inside the
box, like a radiating system. The solution is given in
Fig.~\ref{fig:Onde3}. The same remarks as before apply, outgoing waves
are correctly implemented and well reproduced.

\begin{figure}
  \centering  
  \includegraphics[width=0.45\textwidth]{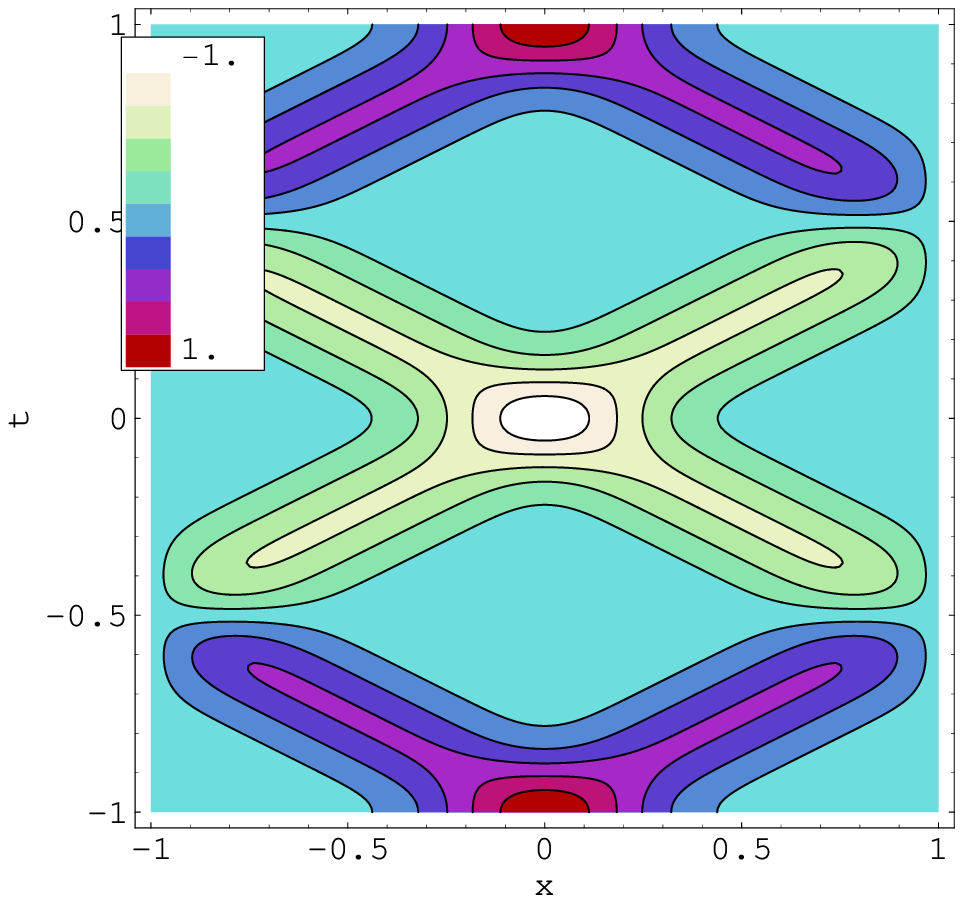}
  \caption{Solution of the wave equation for reflecting boundary
    conditions starting with an initial Gaussian shape.}
  \label{fig:Onde5}
\end{figure}

Finally, we look for reflecting boundary conditions on either one or
two sides. Consider again the initial condition of Fig.~\ref{fig:Onde1}
but with fixed vanishing boundaries. In such a case, the wave is
perfectly reflected and travels back in opposite direction. This is
clearly seen in Fig.~\ref{fig:Onde5}. Note however that this time we
took $v=2$ in order to simulate a full back and forth reflection with
the final state equal to the initial state, allowing easy comparison
and investigation of the accuracy. We know that at the end, we should
retrieve the initial Gaussian shape $u_0(x)$. This is indeed the
case. In addition, the evolution of the relative error between initial
and final state is shown in  Tab.~\ref{tab:OndeErreur5}. It saturates
at a level around $10^{-10}$ for $N_x=N_t=129$.

\begin{figure}
  \centering
  \includegraphics[width=0.45\textwidth]{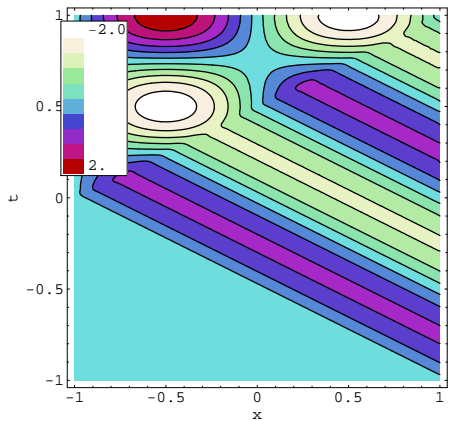}
  \caption{Solution of the wave equation with fixed left boundary
    condition and incoming wave from the right. After a while, at the
    end of the run, a stationary wave settles down with crests and
    nodes.}
\label{fig:Onde7}
\end{figure}

\begin{table}
  \centering
  \begin{tabular}{ccc}
    \hline
    $N_x$ & $N_t$ & error \\
    \hline
    \hline
    5 & 5 & 6.093E-01 \\
    9 & 9 & 1.116E+00 \\
    17 & 17 & 2.170E-01 \\
    33 & 33 & 2.416E-02 \\
    65 & 65 & 8.525E-07 \\
    129 & 129 & 9.858E-11 \\
    \hline    
  \end{tabular}
  \caption{Convergence properties for the wave equation for reflecting
    boundary conditions, fig~\ref{fig:Onde5}. The error is plotted
    against the number of discretisation points, being the same in $x$
    and $t$.}
  \label{tab:OndeErreur5}
\end{table}
Last, we use again vanishing initial conditions with right incoming
sinusoidal wave but left reflecting boundary instead of outgoing wave.
In this case, ingoing and outgoing waves overlap to form a stationary
pattern with crests and nodes as seen in Fig~\ref{fig:Onde7}, it is
equal to $\sin(\pi\,x)$. It is instructive to compare this plot with
Fig~\ref{fig:Onde2}. Their solutions are equal until the signal
reaches the left wall where reflection starts (within a factor two in
timescale because the speeds are different $v=1$ against $v=2$).

\subsection{2D diffusion equation}

We close the tests by looking for a two space dimension problem
symbolize by a diffusion process as described before. For the sake of
simplicity, we only look for uniform diffusion coefficient within the
computational domain, namely $\eta(x,y,t) = \textrm{cste}$, the reason
being that analytical solutions are easily derived. Obviously, the
algorithm can handle non uniform diffusion coefficients without any
modification. 

The initial profile is given by a double sinusoidal profile such that
\begin{equation}
  \label{eq:Diffusion_0}
  u_0(x,y) = \sin (\pi \, x) \, \sin(\pi\,y)
\end{equation}
The exact solution is then given by
\begin{equation}
  \label{eq:Erreur_Diffusion}
  u_{\rm an}(x,y,t) = \sin (\pi \, x) \, \sin(\pi\,y) \, e^{-2\, \eta\,\pi^2 \, t}
\end{equation}
The error is shown in Tab.~\ref{tab:Diffusion_Erreur}. Due to memory
overflow, we could not perform resolutions better than
$33\times33\times33$. Nevertheless, already with a moderate
$17\times17\times17$ resolution, we got an error of only $10^{-13}$.
As expected, the exponential decrease of the error with increasing
number of collocation points is recovered, spectral accuracy is
achieved.
\begin{table}
  \centering
  \begin{tabular}{ccc}
    \hline
    $N_x,N_y$ & $N_t$ & error \\
    \hline
    \hline
    5 & 5 & 6.602E-03 \\
    9 & 9 & 4.194E-06 \\
    17 & 17 & 1.065E-13 \\
    33 & 33 & 5.300E-13 \\
    \hline    
  \end{tabular}
  \caption{Convergence properties for the diffusion equation. The error is plotted
    against the number of discretisation points, being the same in $x$
    and $t$.}
  \label{tab:Diffusion_Erreur}
\end{table}
Including a spatially varying diffusion coefficient is straightforward.

\section{Conclusion\label{sec:Conclusion}}

In this paper, we demonstrated the ease to implement a numerical
algorithm with fully spectral methods in both space and time. We
applied it to three important classes of problems which are the
advection, diffusion and wave equations. The great advantage of our
method is that the treatment is symmetric with respect to space and
time variable.  This reflects in the imposition of the boundary
conditions.  We showed examples for the linear advection, diffusion
and wave equations and recover spectral accuracy as expected.
Different methods were also employed, like penalty, tau and Galerkin
collocation.

The main drawback of the method is the necessity to invert a large
sparse matrix. This limits drastically the number of discretisation
points as well as the number of spatial dimensions.  For multi space
dimensional purposes, the required memory quickly overwhelms the
available capacities of any computer. Moreover the time spend to
invert the matrix becomes prohibitive.  In future work, it is
desirable to adapt the algorithm and put some effort to come over this
difficulty by employing iterative methods for solving the corresponding
large algebraic linear system.

Extension to cylindrical and spherical geometry are also planed, as
those coordinate systems are better adapted to our astrophysical
interests, for instance accretion disks or structure of stars.  Thanks
to the multiplication matrices~$\mathcal{P}$ introduced in this paper,
the discretised version of any linear differential operator is derived
easily. However, outgoing wave boundary conditions in arbitrary
geometry need a more careful treatment. This is left for future work.

The preliminary study presented in this paper was aimed to demonstrate
that fully spectral method are easy to design even for two space
dimensions. We hope that it will serve as a starting point to those
willing to use such techniques for their own particular applications.


\end{document}